\documentstyle[aps,prb,twocolumn,floats,psfig]{revtex}

\begin{document}

\bibliographystyle{prsty}

\wideabs{

\begin{flushleft}
{\small \em submitted to}\\
{\small 
PHYSICAL REVIEW B 
\hfill
VOLUME {\normalsize XX}, 
NUMBER {\normalsize XX} $\qquad\qquad$
\hfill 
MONTH {\normalsize XX}
}
\end{flushleft}

\title{ 
Tunneling of a large spin via hyperfine interactions
}

\author{
D.  A.  Garanin \cite{e-gar} , E. M. Chudnovsky \cite{e-chu}, and R. Schilling \cite{e-sch} 
}

\address{\cite{e-gar}  
Max-Planck-Institut f\"ur Physik komplexer Systeme, N\"othnitzer Strasse 38,
D-01187 Dresden, Germany }

\address{ \cite{e-chu} 
Department of Physics and Astronomy, City University of New York--Lehman College, \\
Bedford Park Boulevard West, Bronx, New York 10468-1589 }

\address{ \cite{e-sch}
Institut f\"ur Physik, Johannes Gutenberg-Universit\"at 
Mainz, Staudingerweg 7, D-6500 Mainz, Germany \\
}

\date{Received 4 November 1999}
\maketitle


\begin{abstract}
We consider a large spin ${\bf S}$ in the magnetic field 
parallel to the uniaxial crystal field, interacting with $N\gg 1$
nuclear spins ${\bf I}_{i}$ via Hamiltonian 
${\cal{H}}=-DS_{z}^{2}-H_{z}S_{z}+
A{\bf S}{\cdot}\sum_{i=1}^N {\bf I}_i$ with $A\ll D$, at temperature $T$.  
Tunneling splittings and the selection rules for the resonant values 
of $H_{z}$ are obtained perturbatively.  The quantum coherence exists 
at $T\ll ASI$ while at $T \gtrsim ASI$ the coherence is destroyed and the 
relaxation of ${\bf S}$ is described by a stretched dependence which can be 
close to $\log t$ under certain conditions. 
 Relevance to Mn$_{12}$ acetate is discussed. 
\end{abstract}     
\smallskip
\begin{flushleft}
\end{flushleft}
}

The effect of nuclear spins on tunneling of a large spin ${\bf S}$
has been the subject of intensive research in the last decade. 
\cite{Garg-93,PS-93,Garg-95,PS-96,Villain,TPS-97}
  The interest to this problem is two-fold. 
Firstly, nuclear spins are likely to significantly affect tunneling
and coherence.  
Secondly, they represent a non-conventional dissipative
environment that cannot be treated by the Caldeira-Leggett method. 
\cite{CL}
Models studied to date 
\cite{Garg-93,PS-93,Garg-95,PS-96,Villain,TPS-97}
assume that tunneling of ${\bf S}$ is induced
by those terms in the Hamiltonian, e.g.,  transverse field or transverse
anisotropy, which do not commute with, e.g., a longitudinal anisotropy term. 
\cite{KS,ES,HS,CG,Zaslavsky,Garanin}  
In that case nuclear spins interfere with tunneling by producing a bias
field on the spin ${\bf S}$ and driving it off resonance. 
 The rate of 
tunneling of ${\bf S}$ then decreases by a statistical factor
that reflects the probability of the environmental spins arranging 
such that $\sum_{i=1}^N {\bf I}_i=0$. \cite{Garg-93,PS-93}  

In this paper we study a different situation when tunneling is 
actually induced by the interaction of ${\bf S}$ with nuclear spins.  
This situation may be relevant to Mn$_{12}$ acetate 
\cite{Sessoli,Friedman,Mirebeau,Wernsdorfer} where the 
crystal-field terms responsible for tunneling are very small. 
We consider the following Hamiltonian
%
%
\begin{equation}\label{Ham}
\hat H = - D S_z^2 -H_z S_z + A \bbox{\rm S \cdot I}_{\rm tot},
\qquad {\bf I}_{\rm tot} = \sum_{i=1}^N {\bf I}_i,
\end{equation}
where the electronic spin $S\gg 1$, ${\bf I}_{\rm tot}$ is the sum of $N\gg 1$ nuclear spins 
with magnitude $I$, and $A\ll D$ is an effective hyperfine constant.  
As an illustration we will keep in mind the example of Mn$_{12}$
with $S=10$, $I=5/2$, $N=12$, $D=0. 6~$K, and $A=2$~mK.  
We shall assume that the nuclear time $T_{1}$ needed for ${\bf I}_{i}$
to flip is larger than the tunneling time for ${\bf S}$.  
The Hamiltonian conserves the magnitude $I_{\rm tot}$ of ${\bf I}_{\rm tot}$, thus dynamically 
it describes a two-spin problem within each subspace of fixed $I_{\rm tot}$. 
On the other hand, for an assembly of Mn$_{12}$ molecules there is a distribution 
over the values of  $I_{\rm tot}$, and all results should be averaged over this distribution. 
Contrary to previous approaches to problems with nuclear spins, we 
shall solve the above-formulated problem rigorously, using the 
smallness of $A$ in comparison with $D$. 

The longitudinal part of the hyperfine interaction, 
$A S_z I_{{\rm tot},z}$ splits the energy levels of the electronic spin which become
%
\begin{equation}\label{Levels}
\varepsilon_{m,m_I} = - D m^2 - H_z m + A m m_I,
\end{equation}
where $m$ and $m_I$ are the projections of ${\bf S}$ and ${\bf I}_{\rm tot}$ on the $z$ axis. 
The transverse part of the hyperfine interaction, 
$(A/2)(S_+ I_{{\rm tot},-} + S_- I_{{\rm tot},+} )$, yields only small corrections 
to the energy levels in the wells in the second order of the perturbation theory which 
are of order $A^2/D$ and can be neglected. 
On the other hand, the transverse spin terms may cause transitions between degenerate
energy levels on different sides of the barrier. 

In contrast to the problem of spin tunneling in the transverse field,
here the projection of the total spin of the system $\bbox{\rm S + I}_{\rm tot}$ 
on the $z$ axis is conserved. 
The value of the bias field $H_z$ for which resonant transitions occur,
is defined by the conservation laws
%
\begin{equation}\label{Conservation}
\varepsilon_{m,m_I} = \varepsilon_{m',m'_I}, \qquad
m + m_I = m' + m'_I
\end{equation}
and can be parametrized by the two quantum numbers $k,k_I$
%
\begin{equation}\label{ResField}
H_z(k,k_I) = Dk + Ak_I,
\qquad 
\end{equation}
where $k$ and $k_I$ satisfy
%
\begin{eqnarray}\label{Relations}
&&
m' = -m -k,  \qquad  m_I = m' + k_I
\nonumber\\
&&
m'_I = m-m'+m_I= m+k_I. 
\end{eqnarray}
Here, by definition, $m<0$ (see Fig.\ 1 of Ref.\ \onlinecite{garchu97} for an illustration of
electronic resonances).
Whereas the number $k$ is integer and labels the main electronic resonance, 
the number $k_I$ may be integer or half integer and describes its hyperfine splitting. 
For the zero-field resonance, $k=k_I=0$, the projection of the total
spin of the system on the $z$ axis is zero for both initial and final states. 
This only takes place, however, if both $S$ and $I_{\rm tot}$ are integer or
half integer. 
If one of them is integer and another half integer, there is no resonance at zero field.

For a fixed value of $I_{\rm tot}$, the possible values of $k_I$ are defined by the conditions 
$-I_{\rm tot} \leq m_I, m'_I \leq I_{\rm tot}$ and satisfy
%
\begin{equation}\label{kIBounds}
-I_{\rm tot} - m = -I_{\rm tot} +m'+k \leq k_I \leq I_{\rm tot} - m'. 
\end{equation} 
One can see that close to the top of the barrier, where $m$ and $m'$ are close to each other,
the number of values of $k_I$ is slightly less then $2I_{\rm tot} + 1$. 
Since $I_{\rm tot} \leq NI$, the maximal number of these resonances is about $2NI + 1$.  
On the contrary, for the unbiased ground-state resonance ($m=-S$, $m'=S$)
there are only $2(I_{\rm tot}-S) +1$ values of $k_I$. 
For systems with $NI < S$, there are no resonant transitions between ground-state
levels. 
 
\begin{figure}[t]
\unitlength1cm
\begin{picture}(11,6)
\centerline{\psfig{file=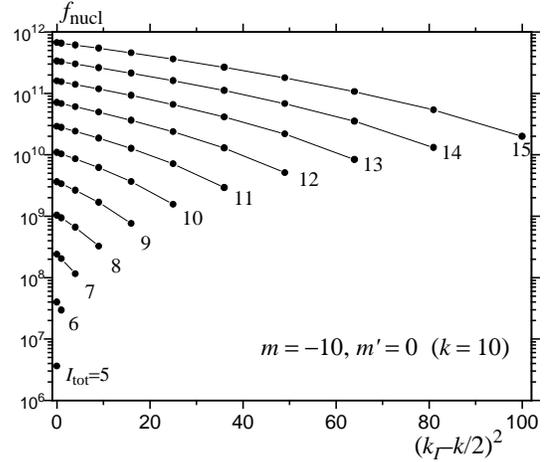,angle=-90,width=8cm}}
\end{picture}
\caption{ \label{nuc_fki}
Dependence of the level splitting $\Delta\varepsilon \propto f_{\rm nucl}$ 
[see Eq.\ (\protect\ref{fNucl})] on the nuclear
quantum number $k_I$ in Eq.\ (\protect\ref{ResField}).
It approaches a Gaussian for high spin values.
}
\end{figure}

Let us consider now the tunneling splitting $\Delta \varepsilon$ between the resonant levels. 
Since the hyperfine constant $A$ is much smaller than the anisotropy $D$,
it can be calculated with the help of the high-order perturbation theory. 
\cite{KS,Garanin,aubflaklaolb96,garchu97}  
One has 
%
\begin{eqnarray}\label{chain}
&&
\Delta \varepsilon_{m,m_I;m',m'_I} = 2
\langle m,m_I | \hat H | m+1, m_I-1 \rangle 
\nonumber\\
&&
\times \frac{1}{ \varepsilon_{m+1, m_I-1} - \varepsilon_{m,m_I} }
\langle m+1, m_I-1 | \hat H | m+2, m_I-2 \rangle
\nonumber\\
&&
\ldots \frac{1}{ \varepsilon_{m'-1, m'_I+1} - \varepsilon_{m,m_I} }
\langle m'-1, m'_I+1 | \hat H | m', m'_I \rangle. 
\end{eqnarray}
Performing products in this formula, one arrives at the final result
\begin{eqnarray}\label{Split}
&&
\Delta \varepsilon_{m,m_I;m',m'_I} = 
\frac{2(D+A)}{[(m'-m-1)!]^2}
\left( \frac{ A }{ 2(D+A) } \right)^{m'-m}
\nonumber\\
&&
\qquad
\times 
\sqrt{\frac{ (S+m')! (S-m)! }{ (S-m')! (S+m)! } 
\frac{ (I_{\rm tot}-m'_I)! (I_{\rm tot}+m_I)! }{ (I_{\rm tot}+m'_I)! (I_{\rm tot}-m_I)! }. 
}
\end{eqnarray}
The part of this expression depending on the state of the nuclear 
subsystem,
%
\begin{equation}\label{fNucl}
f_{\rm nucl}(I_{\rm tot}, k_I) = 
\sqrt{
\frac{ (I_{\rm tot}-m-k_I)! (I_{\rm tot}+m'+k_I)! }{ (I_{\rm tot}+m+k_I)! (I_{\rm tot}-m'-k_I)! }
},
\end{equation} 
has its maximum at $k_I$ in the middle of its interval,
$k_I=-(m+m')/2=k/2$ [see Eq.\ (\ref{kIBounds})], and the minimal value at the borders
 of this interval,  $k_I=-I_{\rm tot} -m$
and $k_I=I_{\rm tot}-m'$. 
If the number of allowed values of $k_I$ is much greater than one,
$f_{\rm nucl}$ approaches a Gaussian (see Fig.\ \ref{nuc_fki}).

For a given resonance $(k,k_I)$ the splitting depends, 
apart from the initial and final electronic states, $m$ and $m'$, 
also on the value of $I_{\rm tot}$. 
Inverting Eq.\ (\ref{kIBounds}), one obtains the range of possible 
values of $I_{\rm tot}$ for a given resonance
%
\begin{equation}\label{ItotRange}
\max\{m'+k_I, -m-k_I, (m'-m)/2\} \leq I_{\rm tot} \leq NI. 
\end{equation} 
The nuclear quantum number $k_I$ itself changes in the range
$-NI -m \leq k_I \leq NI - m'$, as follows from Eq.\ (\ref{kIBounds}) if one replaces 
$I_{\rm tot}$ by its maximal value $NI$. 
According to Eq.\ (\ref{Relations}), one has $m'_I = - NI$ on the left border and 
$m_I = NI$ on the right border of this interval.
For these border values of $k_I$, there is obviously only one possible 
value $I_{\rm tot}=NI$ satisfying Eq.\ (\ref{ItotRange}). 
On the other hand, for $k_I=k/2$ the number of values of  
$I_{\rm tot}$ attains its maximal value of about $NI - m' - k/2$. 
In addition, each value of $I_{\rm tot}$ can be built up of individual nuclear spins 
in a number of different ways, if $I_{\rm tot}< NI$. 
The latter shows that the hyperfine resonances near the center of the splitted electronic
resonance, $k_I=k/2$, are much stronger than
those for $k_I$ away from the center. 
It is, however, difficult to parametrize the strength of these hyperfine lines since the
resonance dynamics is rather complicated, as will be seen shortly. 

The dependence of the level splitting on $I_{\rm tot}$ for a given hyperfine resonance
results in the decoherence of tunneling. 
If electronic spins are prepared in the state $m$, then only molecules with 
$m_I$ satisfying the resonance condition of 
Eq.\ (\ref{ResField}) will
take part in the process. 
Among those with a given $m_I$, there are molecules with different possible 
values of $I_{\rm tot}$, which will show oscillations between the degenerate states with different frequencies.  
If in the initial state nuclear spins are in equilibrium, the probability 
of finding the electronic spin in the initial state $m$ depends on time according to
%
\begin{equation}\label{pOscill}
p_m(t) = 1 - \frac{ e^{-Amm_I/T} }{ {\cal Z}_I^N }
\sum_{I_{\rm tot}} {\cal N}(I_{\rm tot})
[1 - \cos(\Delta \varepsilon t)],
\end{equation} 
where ${\cal Z}_I = \sinh\{[1+1/(2I)]\xi\}/\sinh[\xi/(2I)]$ with $\xi=AmI/T$ is the 
partition function of an individual nuclear spin, $I_{\rm tot}$ satisfies
Eq.\ (\ref{ItotRange}), 
$\Delta \varepsilon=\Delta \varepsilon_{m,m_I;m',m'_I}(I_{\rm tot})$ is given
by Eq.\ (\ref{Split}),  and ${\cal N}(I_{\rm tot})$ is the number of ways
 to combine $I_{\rm tot}$. 
For temperatures larger than the characteristic nuclear temperature, $\xi \ll 1$, the statistical 
factor in front of the sum simplifies to $(2I+1)^{-N}$. 
Under typical conditions, only a small part of the molecules have a given value of
$m_I$ and thus take part in the
resonant tunneling, thus the dynamical term in Eq.\ (\ref{pOscill}) is much
smaller than one. 
This is a mechanism of tunneling reduction due to nuclear spins pointed out in 
Refs.\ \onlinecite{Garg-93} and \onlinecite{PS-93}. 
The rest of molecules which do not have the required value of $m_I$ for a given resonance
are ``frozen in", and they should wait for a longer time $T_1$ required for the nuclear spins to relax.
The study of the latter is beyond the scope of this paper.

\begin{figure}[t]
\unitlength1cm
\begin{picture}(11,6)
\centerline{\psfig{file=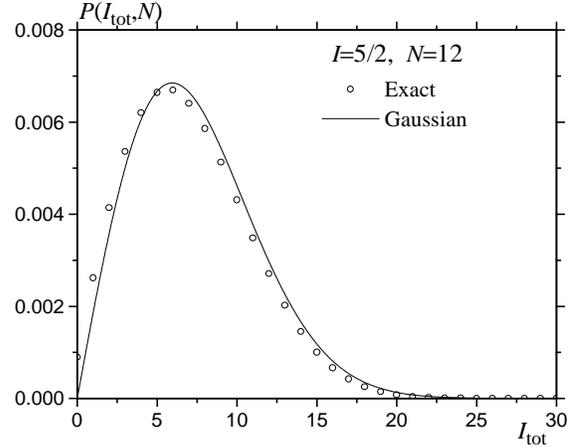,angle=-90,width=8cm}}
\end{picture}
\caption{ \label{nuc_pi}
The normalized number of realizations of the total nuclear spin $I_{\rm tot}$, 
$P(I_{\rm tot})\equiv {\cal N}(I_{\rm tot})/(2I+1)^N$: exact and Gaussian, see 
Eq.\ (\protect\ref{NAsymp}).
}
\end{figure}

The number of realizations ${\cal N}(I_{\rm tot})$ in Eq.\ (\ref{pOscill})  can be computed recurrently. 
If the total spin of a system of $N$ nuclei is $I_{\rm tot}$, the total
spin of its subsystem of $N-1$ nuclei $I'_{\rm tot}$ assumes the values
$|I_{\rm tot}-I| \leq I'_{\rm tot} \leq \min\{I_{\rm tot}+I, (N-1)I\}$. 
Thus for the number of realizations ${\cal N}(I_{\rm tot},N)$ one can write. 
%
\begin{equation}\label{Recurr}
{\cal N}(I_{\rm tot},N) = \sum_{I'_{\rm tot}=|I_{\rm tot}-I|}
^{\min\{I_{\rm tot}+I, (N-1)I\}} {\cal N}(I'_{\rm tot},N-1). 
\end{equation} 
The initial condition for this recurrence relation is 
${\cal N}(I'_{\rm tot},2) = 1$ for $0 \leq I'_{\rm tot} \leq 2I$.  
The quantity ${\cal N}(I_{\rm tot})$ obeys the normalization condition
%
\begin{equation}\label{NNorm}
\sum_{I_{\rm tot}= {\rm frac}(NI)}^{NI} (2I_{\rm tot}+1) 
{\cal N}(I_{\rm tot}) = (2I+1)^{N}. 
\end{equation} 
For $NI \gg 1$, the quantity 
$(2I_{\rm tot}+1){\cal N}(I_{\rm tot})/(2I+1)^{N}$ is the high-temperature
distribution function of the magnitude of $I_{\rm tot}$ and it is well
approximated by $4\pi I_{\rm tot}^2 F({\bf I}_{\rm tot})$, where 
$F({\bf I}_{\rm tot})$ is a normalized Gaussian function with respect to the three
components of ${\bf I}_{\rm tot}$. \cite{garchu97} 
Thus one has the asymptotic form for $NI \gg 1$
%
\begin{equation}\label{NAsymp}
P(I_{\rm tot}) \equiv \frac{ {\cal N}(I_{\rm tot})}{ (2I+1)^{N}}
\cong \frac{ 2\pi I_{\rm tot}  }{ (2\pi\sigma_I)^{3/2} }
\exp\left( - \frac{ I_{\rm tot}^2 }{ 2\sigma_I } \right). 
\end{equation}
where $\sigma_I=(N/3) I(I+1)$.  
It has a maximum at $I_{\rm tot} = \sqrt{\sigma}$ which is about 6 for Mn$_{12}$ 
($I=5/2$, $N=12$). 
Fig.\ \ref{nuc_pi} shows an agreement between the exactly computed 
$P(I_{\rm tot})$ and its Gaussian approximation for $I=5/2$ and $N=12$. 
This agreement improves for higher values of $NI$.

\begin{figure}[t]
\unitlength1cm
\begin{picture}(11,6)
\centerline{\psfig{file=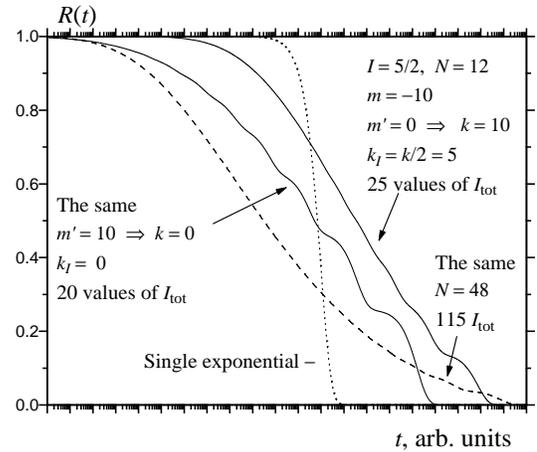,angle=-90,width=8cm}}
\end{picture}
\caption{ \label{nuc_rel}
Stretched relaxation curve, Eq.\ (\protect\ref{Relax}), for the tunneling resonances 
due to nuclear spins.
The sets of parameters labeled by ``The same'' are variations of the basic set of parameters
in the upper right corner.
}
\end{figure}

Due to the summation over $I_{\rm tot}$ in Eq.\ (\ref{pOscill}) and the 
dependence of the splitting $\Delta\varepsilon$ on $I_{\rm tot}$, the coherence of tunneling 
is destroyed.  
On the other hand, tunneling oscillations are damped due to the finite 
linewidth of the electronic levels $\gamma_m$. 
In the overdamped case $\gamma_m \gg \Delta\varepsilon$ the relaxation of the initially prepared state $m$ 
to the matching state $m'$ for $\xi \equiv AmI/T \ll 1$ is described by a sum of exponentials
 (cf. Ref.\ \onlinecite{garchu97})
%
\begin{equation}\label{pRelax}
p_m(t) = 1 -
\sum_{I_{\rm tot}} P(I_{\rm tot})
\frac 12 [1-\exp(-\Gamma_{m,m',k_I,I_{\rm tot}}t)],
\end{equation} 
where $I_{\rm tot}$ satisfies Eq.\ (\ref{ItotRange}),
%
\begin{equation}\label{GammaItot}
\Gamma_{m,m',k_I,I_{\rm tot}} = 
\frac{(\Delta\varepsilon_{m,m',k_I})^2 \gamma_{mm'} }
{ \gamma_{mm'}^2 + (\varepsilon_{m,m_I}-\varepsilon_{m',m'_I})^2 },
\end{equation}
$\gamma_{mm'} \equiv \gamma_m + \gamma_{m'}$, and
$\varepsilon_{m,m_I}-\varepsilon_{m',m'_I}=(m-m')(H-H_{k,k_I})$. 
As we have pointed out above, the process described by Eq.\ (\ref{pRelax}) does not lead to
the full relaxation since only a small fraction of the systems possesses
the required value of the projection $m_I=m'+k_I$ of the total nuclear spin. 
The probability $p_m$ decreases due to this process by 
%
\begin{eqnarray}\label{GSumPItot}
&&
\Delta p_m = \frac 12 \sum_{I_{\rm tot}} P(I_{\rm tot}) \cong \frac 12
\int\limits_{I_{\rm tot, min}}^\infty P(I_{\rm tot}) dI_{\rm tot}
\nonumber\\
&&\qquad
{} = \frac 12 \frac{1}{ \sqrt{2\pi\sigma_I}}
\exp\left( - \frac{ I_{\rm tot, min}^2 }{ 2\sigma_I } \right),
\end{eqnarray}
where Eq.\ (\ref{NAsymp}) was used and $I_{\rm tot, min}$ is defined by  
Eq.\ (\ref{ItotRange}). 
In particular, for the central hyperfine line of an electronic resonance, $k_I=k/2$, one has
$I_{\rm tot, min}=(m'-m)/2$.
For resonances between low-lying electronic levels this value is big and hence $\Delta p_m$ is
small.
An extreme case is the resonance between $m=-S$ and $m'=S$ for which $\Delta p_m=0.008$ for
the Mn$_{12}$ set of parameters.

One should note that the relaxation described by 
Eq.\ (\ref{pRelax}) is not a single exponential but a sum of different 
exponentials with a faster rate and smaller amplitude 
(larger $I_{\rm tot}$) and slower rate and larger amplitude 
(smaller $I_{\rm tot}$).  
To illustrate this fact we have plotted in Fig.\ \ref{nuc_rel} 
the relaxation function
%
\begin{equation}\label{Relax}
R(\tilde t) = \frac{ \sum_{I_{\rm tot}} P(I_{\rm tot})
\exp[-f_{\rm nucl}^2(I_{\rm tot}, k_I)\tilde t] }
{\sum_{I_{\rm tot}} P(I_{\rm tot}) },
\end{equation} 
where $f_{\rm nucl}$ is given by Eq.\ (\ref{fNucl}) and $\tilde t$ is the scaled time
including all factors which do not depend on $I_{\rm tot}$ and $k_I$. 
One can see that for the chosen values of the parameters, the relaxation function
$R(\tilde t)$ is stretched for about 14 decades, whereas a single exponential
practically decays within two decades in time. 
To analytically clarify the large-time behavior of $R(\tilde t)$ one can apply 
 the Stirling formula to Eq.\ (\ref{fNucl}) to obtain
\begin{equation}
\label{Stirling}
f_{\rm nucl}^2 \cong (e I_{\rm tot})^{2(m'-m)}
\frac{\left[1-\left(\frac{m'+k_I}{I_{\rm tot}}\right)^2\right]^{m'+k_I} }
{\left[1-\left(\frac{m+k_I}{I_{\rm tot}}\right)^2\right]^{m+k_I}}.
\end{equation}
Here the second factor is only weakly dependent on $I_{\rm tot}$. 
For large $m'-m$, the dominant  dependence on $I_{\rm tot}$ is given by the first factor. 
Substituting Eq.\  (\ref{Stirling}) into Eq.\  (\ref{Relax}), 
replacing the sum by an integral and introducing
$\tilde{I}_{\rm tot} = I_{\rm tot}\;\tilde{t}^{\frac{1}{2(m'-m)}}$
as a new integration variable, one obtains from Eqs.\ (\ref{NAsymp}) and (\ref{Relax})
the large-time behavior
\begin{equation}
\label{power}
R(\tilde{t}) \propto \tilde{t}^{-\frac{1}{(m'-m)}}.
\end{equation} 
Since $m'-m$ can be large, e.g., about ten, the exponent can be rather small. 
Therefore Eq.\ (\ref{power}) can be approximated by :
\begin{equation}\label{log}
R(\tilde{t}) \propto 1- \frac{1}{m'-m} \ln\tilde{t},
\end{equation}
which shows that the relaxation, although being a power law, may be well approximated by a
logarithmic time dependence.
We remind the reader that $m'-m$ is always positive.

\begin{figure}[t]
\unitlength1cm
\begin{picture}(11,6.5)
\centerline{\psfig{file=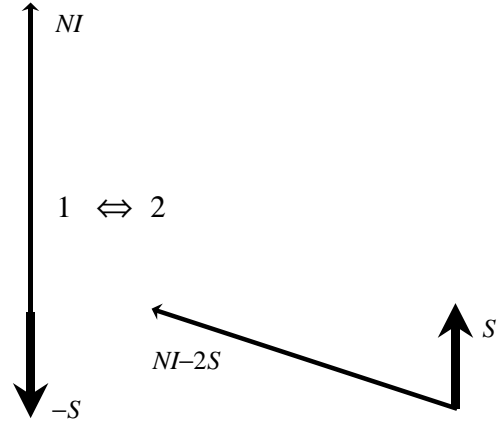,angle=0,width=14cm}}
\end{picture}
\caption{ \label{nuc}
Tunneling between electronic ground states $\pm S$ at zero temperature.  The $z$ projection of the total spin is conserved.  To conserve energy,
the field $H_z=Ak_I$ with $k_I=NI-S$ should be applied [see Eq.\ (\protect\ref{Relations})]. }
\end{figure}

\begin{table}
 \caption{
 Tunneling splittings due to the nuclear spins for the zero-temperature resonances $m=-S$, $m'=S-k$ for the Mn$_{12}$
 set of parameters. }
 \begin{tabular}{ll} 
 $k$ & $\Delta\varepsilon$, K\\ 
\tableline 
 10  &  2.273148E-18\\ 
 11  &  4.679795E-16\\
 12  & 8.019877E-14\\
 13  & 1.126574E-11\\
 14  &  1.268525E-09\\
 15  &  1.107789E-07\\
 16  &  7.129369E-06\\
 17  &  3.102200E-04\\
 18  &  7.706470E-03\\
 19  &  6.928205E-02
 \end{tabular}
 \end{table}

For temperatures much lower than the nuclear temperature, i.e., for $\xi \equiv AmI/T \gg 1$,
the coherence is restored, since the nuclear spins fall in the lowest-energy
 state $m_I=I_{\rm tot}=NI$, and there is only one term in Eq.\ (\ref{pOscill}).
 This term corresponds to $m=-S$, $m'=S-k$, $m_{I}=NI$, and $m'_{I}=NI-2S+k$, i.e., 
 $k_{I}=NI-S+k$ (see Fig.\ \ref{nuc}). 
Consequently, the low-temperature resonances are characterized by 
only one number $k$:
\begin{equation}
H_{z}(k)=A(NI-S)+ (D+A)k.
\end{equation}
In the case of Mn$_{12}$ the corresponding quantum relaxation
due to nuclear spins can become observable at $k>13$, as is
illustrated by Table 1 that lists tunneling splittings for
$10<k<19$.  
Note that for large values of $k$, tunneling due
to hyperfine interactions in Mn$_{12}$ should dominate over the 
effect of other small non-diagonal terms unaccounted for in 
Eq.\  (\ref{Ham}). 

In fact, the simplest model used in this paper cannot be quantitatively accurate for Mn$_{12}$
since each molecule of Mn$_{12}$ contains eight Mn atoms with the spin value 2 and four Mn
atoms with the spin 3/2.
The latter couple ferromagnetically within each group, and the two groups couple
ferrimagnetically with each other to build the total spin $S=10$ (see the details in the
recent Ref.\ \onlinecite{katdobhar99}).
The hyperfine Hamiltonian of a Mn$_{12}$ molecule contains {\em two} hyperfine constants $A$
and $A'$ for each type of Mn atoms.\cite{Villain}
Thus nuclear spins are splitted into two groups which behave dynamically as two effective
``giant'' nuclear spins, if the nuclear relaxation is neglected.
The resulting three-spin model (one electronic and two nuclear spins) is more complicated
than the simplified two-spin model considered above, although tunneling splittings between
different resonant states can be obtained by the same method, cf.\ Eq.\ (\ref{chain}).
The resonant values of the bias field $H_z$ are parametrized by {\em four} quantum numbers
instead of two in Eq.\ (\ref{ResField}) and hence there are much more hyperfine lines for each
electronic resonance.
We do not try to work out this more realistic model here since the calculations should be
rather cumbersome and the accurate values of $A$ and $A'$ are unknown.
This, however, does not affect our qualitative conlusions derived from a simplified model.

In conclusion, we have studied tunneling and relaxation of
a large spin, induced by the hyperfine interactions. 
This model can be relevant to Mn$_{12}$ acetate in the absence of the
transverse field. 
 Our main findings are these.  There must be two distinct temperature
regimes, above and below the hyperfine temperature, $T_{hf}=ASI$,
which for Mn$_{12}$ is about 50~mK.  
At $T \gtrsim T_{hf}$ the magnetic 
relaxation is stretched and close 
to logarithmic due to the contribution of a large number of 
nuclear states.  
At $T \ll T_{hf}$ only one nuclear state contributes
to the magnetic relaxation at each resonant field, and the 
coherence of tunneling is restored. 

This work has been supported by the NSF Grant No.\  DMR-9024250. 

\end{document}